\documentclass[aps,longbibliography,
  twocolumn,amsmath,
  amssymb,nofootinbib, superscriptaddress, preprintnumbers]{revtex4-1}

\usepackage{graphics}      
\usepackage{graphicx}      
\usepackage{longtable}     
\usepackage{url}           
\usepackage{bm}            
\usepackage{xcolor}
\usepackage{mathrsfs}
\usepackage[colorlinks,allcolors=blue]{hyperref}
\usepackage{blindtext}
\usepackage{hyperref}
\usepackage{float}
\usepackage{bbold}
\usepackage{lipsum}
\usepackage[normalem]{ulem}
\usepackage{orcidlink}
\usepackage{academicons}
\usepackage{xcolor}
\usepackage{subfigure}
\usepackage{booktabs}
\usepackage{hyperref}

\newcommand{\orcid}[1]{\href{https://orcid.org/#1}{\textcolor[HTML]{A6CE39}{\aiOrcid}}}

\begin{document}
\preprint{MPP-2023-39}

\title{Applications of Machine Learning to  Detecting Fast Neutrino Flavor Instabilities in Core-Collapse Supernova and  Neutron Star Merger Models}

\newcommand*{\MPP}{\textit{\small{Max-Planck-Institut f\"ur Physik (Werner-Heisenberg-Institut), F\"ohringer Ring 6, 80805 M\"unchen, Germany}}}
\author{Sajad Abbar \orcidlink{0000-0001-8276-997X}   \\  \MPP } 


\begin{abstract}
  Neutrinos propagating in a dense neutrino gas, such as those expected in core-collapse supernovae (CCSNe) and neutron star mergers (NSMs),
can experience  fast flavor conversions on relatively short scales.
This can happen if
   the neutrino electron lepton number ($\nu$ELN) angular distribution
crosses zero in a certain direction.
 Despite this, most of the state-of-the-art  CCSN and NSM simulations do not provide such detailed angular information and instead, supply
only a few moments of the neutrino angular distributions. 
In this study we employ, for the \emph{first} time, a machine learning (ML) approach
to this problem and  show that it can be extremely successful in detecting $\nu$ELN crossings on the basis of its zeroth and first moments.
We observe that an accuracy of $\sim95\%$ can be  achieved by the ML algorithms, which  almost corresponds to the Bayes error rate of our problem.
Considering its remarkable efficiency and agility,
the ML approach provides one with an unprecedented opportunity 
  to evaluate the occurrence of FFCs in CCSN and NSM simulations \emph{on the fly}. We  also provide our ML methodologies  on
\href{https://github.com/sajadabbar/ML-nu_FFI/tree/main}{GitHub}.

 \end{abstract}

\maketitle

\section{Introduction}
Neutrino emission is a major process in 
core-collapse supernova (CCSN) explosions and neutron star mergers (NSMs).
During their propagations in such dense media, neutrinos can experience
 flavor oscillations in a nonlinear and collective manner, due to their coherent forward scatterings with the 
dense background neutrino 
gas~\cite{pantaleone:1992eq, sigl1993general, Pastor:2002we,duan:2006an, duan:2006jv, duan:2010bg, Mirizzi:2015eza, Volpe:2023met}.
Specially, 
 neutrinos can  undergo \emph{fast} flavor
conversions (FFCs)  on scales $\sim G_{\rm{F}}^{-1} n_{\nu}^{-1}$,
which can be much shorter than those expected in vacuum (see, e.g., Refs.~\cite{Sawyer:2005jk, Sawyer:2015dsa,
Chakraborty:2016lct, Izaguirre:2016gsx,Capozzi:2017gqd,Abbar:2018beu,Capozzi:2018clo, Martin:2019gxb, Capozzi:2019lso, Johns:2019izj, Martin:2021xyl, Tamborra:2020cul,  Sigl:2021tmj, Kato:2021cjf,  Morinaga:2021vmc, Nagakura:2021hyb,  Sasaki:2021zld, Padilla-Gay:2021haz, Abbar:2020qpi, Capozzi:2020syn, DelfanAzari:2019epo, Harada:2021ata, 
Padilla-Gay:2022wck, Capozzi:2022dtr, Zaizen:2022cik, Shalgar:2022rjj,  Kato:2022vsu, Zaizen:2022cik,  Bhattacharyya:2020jpj, Wu:2021uvt, Richers:2021nbx, Richers:2021xtf, Dasgupta:2021gfs, Nagakura:2022kic, Ehring:2023lcd}),
with  $n_\nu$ and $G_{\rm{F}}$ being the neutrino number density and the 
Fermi coupling constant, respectively.

FFCs occur \emph{iff} the angular distribution of the neutrino lepton
number changes its sign in some direction.
Assuming that $\nu_x$ and $\bar\nu_x$ have similar  
angular distributions,  FFCs exist provided that  the angular distribution of the
neutrino electron lepton number ($\nu$ELN) defined as,
\begin{equation}
  G(\mathbf{v}) =
  \sqrt2 G_{\mathrm{F}}
  \int_0^\infty  \frac{E_\nu^2 \mathrm{d} E_\nu}{(2\pi)^3}
        [f_{\nu_e}(\mathbf{p})- f_{\bar\nu_e}(\mathbf{p})],
 \label{Eq:G}
\end{equation}
crosses zero at some $\mathbf{v} = \mathbf{v}(\mu,\phi_\nu)$, with $\mu =\cos\theta_\nu$~\cite{Morinaga:2021vmc}. 
Here, $E_\nu$, $\theta_\nu$, and $\phi_\nu$ are the neutrino energy,  
the zenith, and azimuthal angles of the neutrino velocity, respectively, 
and  $f_{\nu}$'s are the neutrino 
occupation numbers.

Though searching for $\nu$ELN crossings requires access to the full
neutrino angular distributions, 
 such detailed angular information is
not available in most of the state-of-the-art  CCSN and NSM simulations,
due to its unbearable computational 
cost. 
Instead, the
neutrino transport is performed  by considering 
a few number of the  moments
of the neutrino angular distributions defined as,
 \begin{equation}
I_n = \int_{-1}^{1} \mathrm{d}\mu\ \mu^n\ \int_0^\infty \int_0^{2\pi} \frac{E_\nu^2 \mathrm{d} E_\nu \mathrm{d} \phi_\nu}{(2\pi)^3} \
        f_{\nu}(\mathbf{p})\footnote{Note that we 
       here only consider the moments in $\mu$. See the discussion in the  DISCUSSION AND OUTLOOK regarding this point.}.
\end{equation}
For instance, in the $M_1$ closure scheme, only the evolution of $I_0$ and $I_1$
are followed directly, to which  $I_2$ and
$I_3$ are  related analytically through  closure relations~\cite{Just:2015fda, Murchikova:2017zsy}.

Despite the fact that   a huge amount  of  information is lost by following
only a few neutrino angular moments, one can still design methods to utilise
these few moments for
assessing the occurrence of FFCs in CCSN and NSM simulations. 
Generally speaking, such methods can be divided into three different categories.
 In the first category, the instability criteria of some specific Fourier modes are considered  to search for 
 $\nu$ELN crossings~\cite{Dasgupta:2018ulw, Johns:2021taz, Johns:2019izj}.
 The second category is based on
  finding a positive polynomial of $\mu$
   for which the weighted integration of $\nu$ELN
   is negative (positive) for $\alpha<1$ ($\alpha>1$)~\cite{Abbar:2020fcl}.
  In the third category, one can effectively find fits to some phenomenological neutrino
  angular distributions given the few available neutrino angular moments~\cite{Nagakura:2021hyb, Johns:2021taz, Richers:2022dqa}. 
  While the first and second categories are mathematically certain in the crossings they
  detect, the third one is very efficient though it can be associated with some uncertainties. 
  On the other hand, though the first category is normally easy to be checked, the second and third ones
  are computationally  expensive and have been  only tested in the post-processing step.
  Note, however, that despite its easy implementation, the first type of methods is not generally very 
  efficient  in detecting  $\nu$ELN crossings because it is solely sensitive to 
  some specific modes, which are not guaranteed to be always unstable.
  
  In this paper, we consider searching for $\nu$ELN crossings 
  as a classification problem and we show that machine learning (ML) algorithms can be successfully employed to
  detect  $\nu$ELN crossings in the CCSN and NSM models. In particular, we argue that the application of ML in this field is of
 the  utmost importance  since it introduces  a method which is both very efficient in detecting the $\nu$ELN crossings 
 and also can be computationally   very cheap  once the ML algorithm is trained. This provides an opportunity for detecting 
   $\nu$ELN crossings  in the CCSN and NSM simulations \emph{on the fly}.
 In the next section, we first discuss 
 how we prepare our training/test datasets and then 
  we introduce the ML algorithms which we use in our calculations and we present their accuracies.
  Finally and before the conclusion section, we  confirm the reliability of our ML algorithms to a realistic  dataset obtained from a simulation of a NSM remnant model.

 \section{Machine Learning Approach}\label{sec:main}
 We have been facing a data revolution during the last almost three decades, due to an 
 unprecedented enhancement in our power in storing and  processing large datasets.
 This has particularly led to an exceptional  opportunity for learning from data, e.g., in the form
 of ML, which is a subcategory  of artificial intelligence where
 one develops and trains algorithms to detect possible patterns in the data. 
 
 In the case of CCSN and NSM physics, ML has been 
 extensively used in the literature, e.g., 
  to evaluate/detect  their gravitational wave signals~\cite{George:2017pmj, Qiu:2022wub, Aveiro:2022bkk, Krastev:2019koe,  Wei:2020sfz, Schafer:2020kor, Mitra:2022nyc, LopezPortilla:2020odz}, 
 to predict the explosion outcome of a CCSN~\cite{Tsang:2022imn}, and to model the turbulence in CCSNe~\cite{Karpov:2022tro}.(See, e.g., Ref.~\cite{Mehta:2018dln} for 
 reviews on ML for physicists. One can also check Abu
Mostafa’s fascinating lectures on ML fundamentals on \href{https://www.youtube.com/watch?v=mbyG85GZ0PI&list=PLD63A284B7615313A} {YouTube}).

In this section, we introduce another  application of ML in this field and we  discuss a number
of  algorithms which  work successfully  in detecting $\nu$ELN crossings  in CCSN and NSM simulations.

 \begin{figure*} [t!]
\centering
\begin{center}
\includegraphics*[width=1.\textwidth, trim= 10 10 10 10, clip]{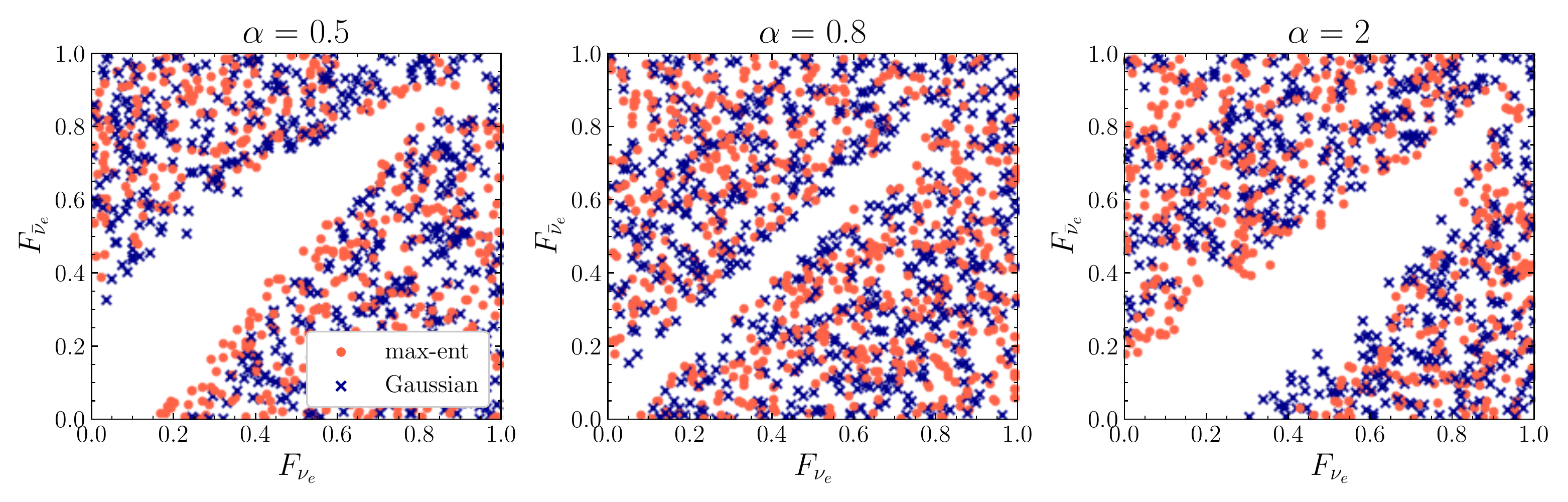}
\end{center}
\caption{The points/crosses show
the $F_{\nu_e} - F_{\bar\nu_e} $'s 
 for which  there is a  crossing in the $\nu$ELN  angular distribution for 
the maximum entropy and Gaussian  distributions, for some values of $\alpha$.
 Note that the crossing (no-crossing) region patterns  are very similar for the two angular 
 distributions. The classifications can be different only at some narrow zones at the boundary 
 of the crossing/no-crossing regions, where the labels get \emph{noisy} meaning that there are different
 labels at the same point.
}
\label{fig:cross}
\end{figure*}

\subsection{Parametric angular distributions}
 ML  requires data to be trained, evaluated, and then tested on.
In order to train and test our ML algorithms, we need a sufficient amount of 
labeled data containing $I_0$ and $I_1$ of $\nu_e$ and $\bar\nu_e$, where the labels
determine whether $\nu$ELN crossing exists or not. 
Note that we here only focus on the first two moments since they are the ones which are normally \emph{directly}
followed in the simulations. 

Though one can, in principle, use the very few available  simulations where the neutrino  angular distributions are directly available, we here 
 train/test our ML algorithms by
 using parametric neutrino angular distributions.
 This choice is quite  justifiable  considering   a number of observations.
 First, one should note that  the moment-based CCSN and NSM simulations use different closure relations.
 This  introduces some variations  which depends on the type of the employed closure relation.
 Therefore, one should already expect a level of uncertainty in this problem once even the angular
 distributions from realistic simulations are used to train the ML algorithms.
 In addition and 
 as indicated in  Fig.~\ref{fig:cross},  reasonable  angular distributions should lead
 to  similar patterns in the parameter space of $I_0$ and $I_1$, regarding the occurrence of  $\nu$ELN crossings.
 This implies that the parametric angular distributions should be able to provide a reliable/acceptable evaluation of the 
 occurrence of $\nu$ELN crossings, given the uncertainties already existing in CCSN and NSM simulations.
 We elaborate more on this later in this section when we test our ML algorithms for detecting $\nu$ELN crossings in
  a time snapshot of a NSM remnant simulation. 
Besides,  considering 
   parametric angular distributions helps one have access to a larger training dataset, hence
reducing the risk of overfitting which could be crucial in this problem. Indeed, using parametric angular distributions  provides one with
a more generic training dataset, which as we see later on is very important in this problem.


We use two parametric neutrino angular distributions which have beed  widely used in the literature.
The first one is the maximum entropy distribution
defined as, 
\begin{equation}
f^{\rm{max-ent}}_\nu(\mu) = \exp[\eta + a\mu],
\end{equation}
where we here consider the $\phi_\nu$-integrated distribution, i.e.,  
\begin{equation}
 f_{\nu}(\mu) =  \int_0^\infty \int_0^{2\pi} \frac{E_\nu^2 \mathrm{d} E_\nu \mathrm{d} \phi_\nu}{(2\pi)^3} 
        f_{\nu}(\mathbf{p}).
\end{equation}
This is a very natural choice for the neutrino angular distribution since
the maximum entropy closure \cite{Cernohorsky:1994yg} is currently very popular  in the  moment-based neutrino
transport methods. 
This  parametric  distribution has been also used to detect $\nu$ELN crossings~\cite{Richers:2022dqa}.
Another angular distribution considered in the literature of FFCs (see, e.g., Refs.~\cite{Wu:2021uvt, Yi:2019hrp})  is the Gaussian distribution defined as,
\begin{equation}
f^{\rm{Gauss}}_\nu(\mu) = A\exp[-\frac{(1-\mu)^2}{a}].
\end{equation}
Note that both of these distributions have a parameter which determines the overall neutrino number density, namely
$\eta$ and $A$, and the other parameter  determining the shape of the distribution, i.e., $a$.

\subsection{Feature engineering}
Thought we here considered $I_0$ and $I_1$ of  $\nu_e$ and $\bar\nu_e$ as the relevant information on its basis one should decide whether
$\nu$ELN crossings exist or not, one should still find the best features on their basis $\nu$ELN crossings can be detected most 
efficiently. Given this, one should first keep in mind that an overall normalisation factor does not affect the occurrence
of  $\nu$ELN crossings. This lead us to consider only $ I^{\bar\nu_e}_0/I^{\nu_e}_0$, instead of $ I^{\bar\nu_e}_0$ and $I^{\nu_e}_0$ independently.
In addition, one can do a similar sort of normalisation for 
$ I^{\bar\nu_e}_1$ and $I^{\nu_e}_1$ given the fact that only the shapes of the angular distributions matter. Having said that,
we consider 
\begin{equation}
\alpha = I^{\bar\nu_e}_0/I^{\nu_e}_0,\  F_{\nu_e} = I^{\nu_e}_1/I^{\nu_e}_0,\ \mathrm{and}\ F_{\bar\nu_e} = I^{\bar\nu_e}_1/I^{\bar\nu_e}_0,
\end{equation}
as the relevant features to be considered in the ML algorithms in this problem.

\subsection{Data preparation}

To cover the $\alpha$ space more efficiently, we consider 84 bins in the range $\alpha = (0.03 - 2.5)$.
Then for each $\alpha$, we generate  700 random points in the $F_{\nu_e}-F_{\bar\nu_e} $ parameter space
for which we find  the maximum entropy and Gaussian angular distributions and then determine whether a $\nu$ELN crossing exists or not. 

In Fig.~\ref{fig:cross}, we show the zones for which there is a  crossing in the $\nu$ELN  angular distribution,
in the $F_{\nu_e} - F_{\bar\nu_e} $ parameter space 
for some values of $\alpha$.
 One should note that the crossing/no-crossing regions are very similar for the two angular 
 distributions, apart from a narrow zone in the boundary  between the crossing/no-crossing
 regions. The fact that there exists a sort of common pattern confirms that this problem can be addressed in the
 framework of ML and also again justifies the application of parametric angular distributions for training
 our ML algorithms.
 
 
In order to train/test the ML algorithms, one should split this  dataset into three independent sets:
i) The \textbf{training} set, which is used to train the ML algorithm, ii) The \textbf{development} set, which is used to find out the hyper-parameters
of the algorithm, and iii) the \textbf{test} set, which provides a measure of the performance of the ML method on new unseen data.

\subsection{Performance metrics}

Before moving on to our ML  algorithms, we would like to comment on different metrics used to evaluate the performance 
of a ML  algorithm. Apart from accuracy which is perhaps the most trivial metric, one can also consider precision,
 recall, and $F_1$ metrics,
\begin{equation}
\begin{split}
&\mathrm{accuracy} = \frac{T_p + T_n}{T_p + T_n + F_p + F_n} \\
&\mathrm{precision} = \frac{T_p}{T_p+F_p} \\
&\mathrm{recall} = \frac{T_p}{T_p+F_n} \\
&F_1 = 2\times \frac{\mathrm{precision} \times \mathrm{recall} }{\mathrm{precision} + \mathrm{recall}},
\end{split}
\end{equation}
with $T(F)_{p(n)}$ denoting True (False) positive (negative) classifications.
An astute reader can see that the precision/recall metric provides information on the reliability/detectability of the positive classification, while 
$F_1$ is their  harmonic mean. In this study, we consider the accuracy as the appropriate metric because we would like to be equally 
sensitive to the existence/absence of the ELN crossings.

 \subsection{ML algorithms}
 In this part, we discuss our results on detecting  $\nu$ELN crossings with the application of  ML algorithms.
 We first focus on  Logistic Regression (\textbf{LR}), which turns out to be the most 
 promising ML algorithm to be used in detecting  $\nu$ELN crossings. Then for illustration purposes, we also discuss 
 our results for a few other algorithms. 
 The accuracies of different algorithms 
are presented in  Table~\ref{tab:tab}.
Most of the methodologies of this part are available on \href{https://github.com/sajadabbar/ML-nu_FFI/tree/main}{GitHub}. 

LR is a classifier which uses the logistic function,
\begin{equation}
\sigma (s)={\frac {1}{1+e^{-s}}},
\end{equation}
on top of a linear one, i.e., $s = \mathbf{x}^T \mathbf{w}$, where $\mathbf{x}$ and $\mathbf{w}$ are
the features and the trained wights, respectively. If $\sigma (s) \geq p_{\mathrm{c}} $ (with $p_{\mathrm{c}}$ being a threshold probability which is here taken to be   $p_{\mathrm{c}} = 0.5$), then the point is classified
into class 1 (crossing) and otherwise, into class 0 (no-crossing). 

\begin{figure} [t]
\centering
\begin{center}
\includegraphics*[width=.45\textwidth, trim= 0 0 30 20, clip]{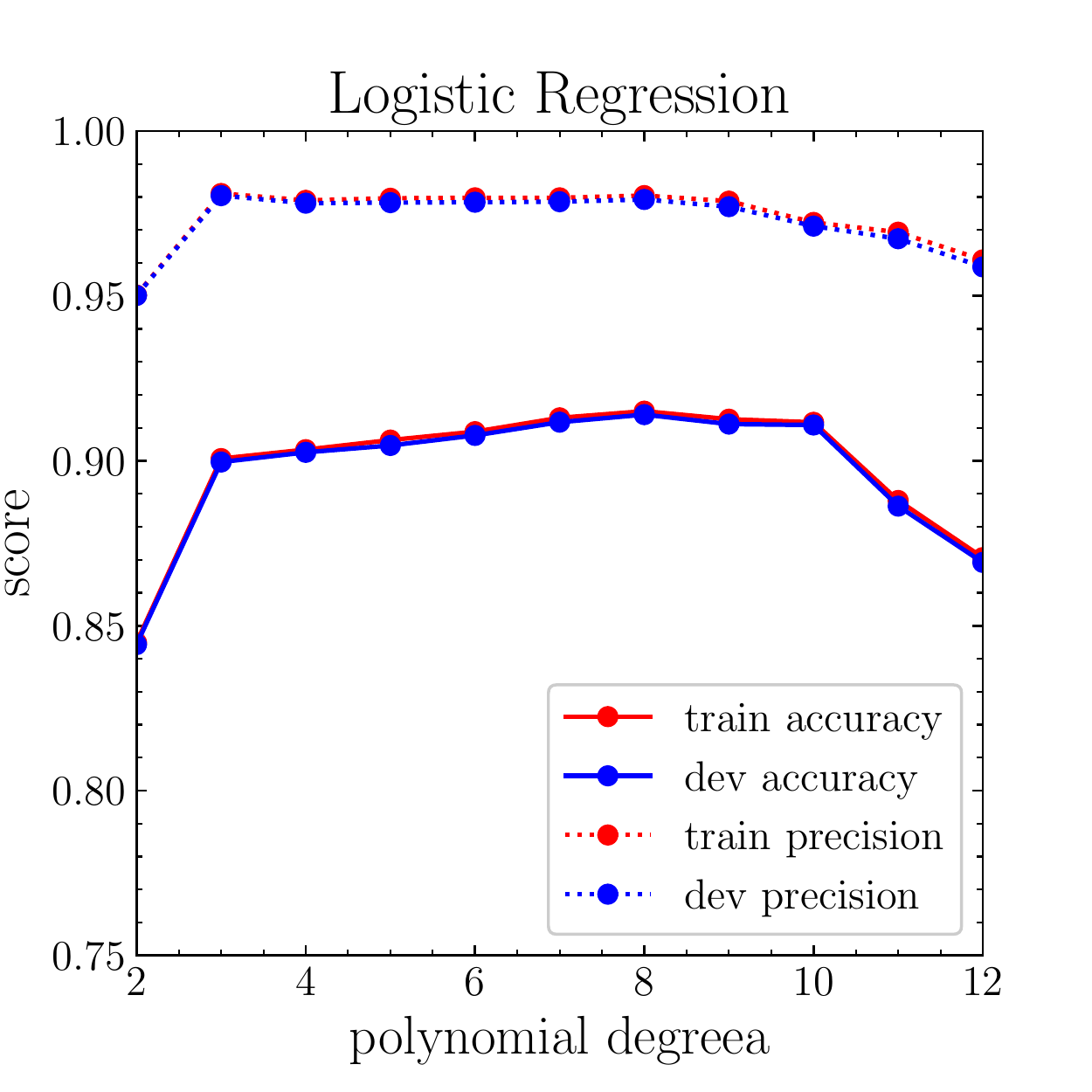}
\end{center}
\caption{
The accuracy and precision of the LR algorithm for the training and development datasets as a function of the 
polynomial degree of the nonlinear transformations. 
The scores reach their maximum at $n\simeq9$, though they are more or less acceptable for $n\gtrsim3$. }
\label{fig:LRdegree}
\end{figure}

\begin{figure} [t]
\centering
\begin{center}
\includegraphics*[width=.45\textwidth, trim= 0 0 30 20, clip]{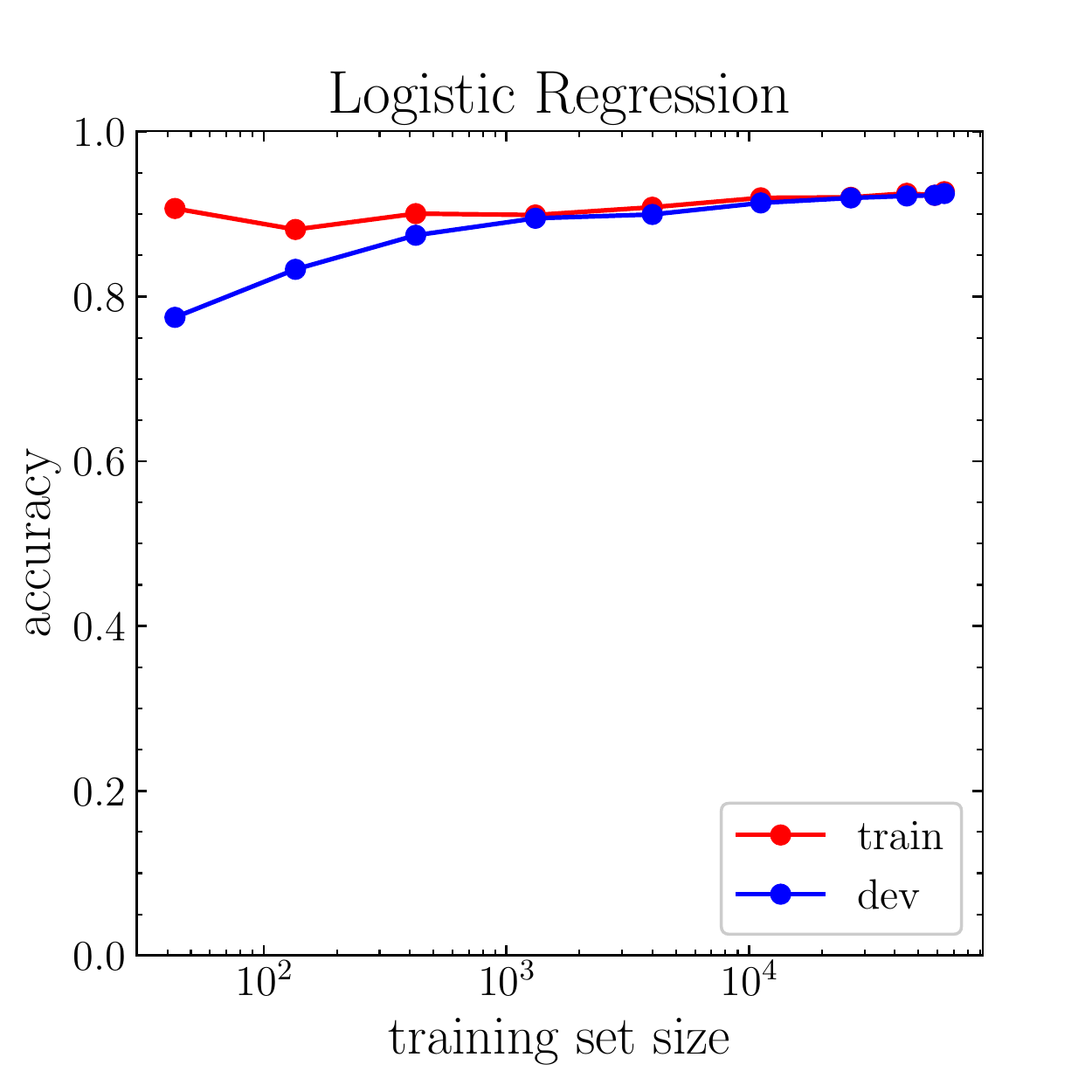}
\end{center}
\caption{
The accuracy of the LR algorithm for the training and development sets. 
Note that the overfitting disappears once at least a few thousand points are considered to train the LR algorithm.}
\label{fig:LRn}
\end{figure}

Though  LR includes the nonlinear logistic function, it is a linear classifier meaning that it can not be used directly
to our problem, which obviously possesses non-linear patterns as presented in Fig.~\ref{fig:cross}.
To address this issue,  one should first make non-linear transformations and generate new features out of the original three
features in  the problem. \textsc{Python sklearn} provides a module which does this, given the polynomial degree of the
nonlinear transformation, which is a hyper-parameter of this algorithm. 

 In Fig.~\ref{fig:LRdegree}, we show the accuracy and precision of the LR algorithm for the training and development datasets. 
 One can easily see that the accuracies reach their maximum at $n\simeq 8-10$, where $n$ is the polynomial degree, though
  we observed that $n=9$ performs a bit better once realistic data are considered (see Sec.\ref{sec:real}).
 Despite this, the metric scores are generally acceptable for $n\geq 3$.

A very interesting characteristic of the LR algorithm is in that it provides a probability interpretation of
the classification. This is possible owing to the fact that the logistic function has a range of $(0,1)$.
One can then write:
\begin{equation}
\begin{split}
P(y=1| \mathbf{x}, \mathbf{w}) &= \sigma( \mathbf{x}^T \mathbf{w}),\\
P(y=0| \mathbf{x}, \mathbf{w}) &= 1 - P(y=1| \mathbf{x}, \mathbf{w}).
\end{split}
\end{equation}

Such an interpretation is indeed very important in our application of ML in detecting $\nu$ELN crossings.
This is because one can then play with the threshold probability, $p_{\mathrm{c}}$,
 to artificially  facilitate/hinder the occurrence of FFCs in the regions of interest. This obviously provides one with a strong tool to measure the
extent of the impact of FFCs on the CCSN and NSM physics.

Regarding the application of ML in detecting  $\nu$ELN crossings in CCSN and NSM simulations, there is another important feature
associated with the  LR algorithm, namely its easy implementation. Indeed once the LR algorithm has been trained and $\mathbf{w}$'s have
been learned, the  LR algorithm can be directly implemented in CCSN and NSM simulation codes  in a quite straightforward manner.

One of the most serious issues in ML is  the overfitting problem, where the performance
of the algorithm in the training set is significantly better than that of the test set, due to the 
unavailability of sufficient amount of data.
This  can lead to a remarkable  variation in the performance once
moving from one dataset to the other.
In Fig.~\ref{fig:LRn}, we show the accuracy of LR as a function of the training set size.
One can see that at least a few thousand data points should be available
to avoid overfitting in this problem. Note that the size  of the dataset is here  determined by
the number of $\alpha$'s multiplied by the number random points   in the  $F_{\nu_e}-F_{\bar\nu_e} $ parameter space.
As we  discuss later on, this number can be very sensitive to the range of $\alpha$
accessible in the training dataset. Therefore if some realistic angular distributions are used for training purposes, one
should make sure that an appropriate  range of $\alpha$ exists in the dataset.

 \begin{figure} [tb!]
\centering
\begin{center}
\includegraphics*[width=.45\textwidth, trim= 0 10 30 20, clip]{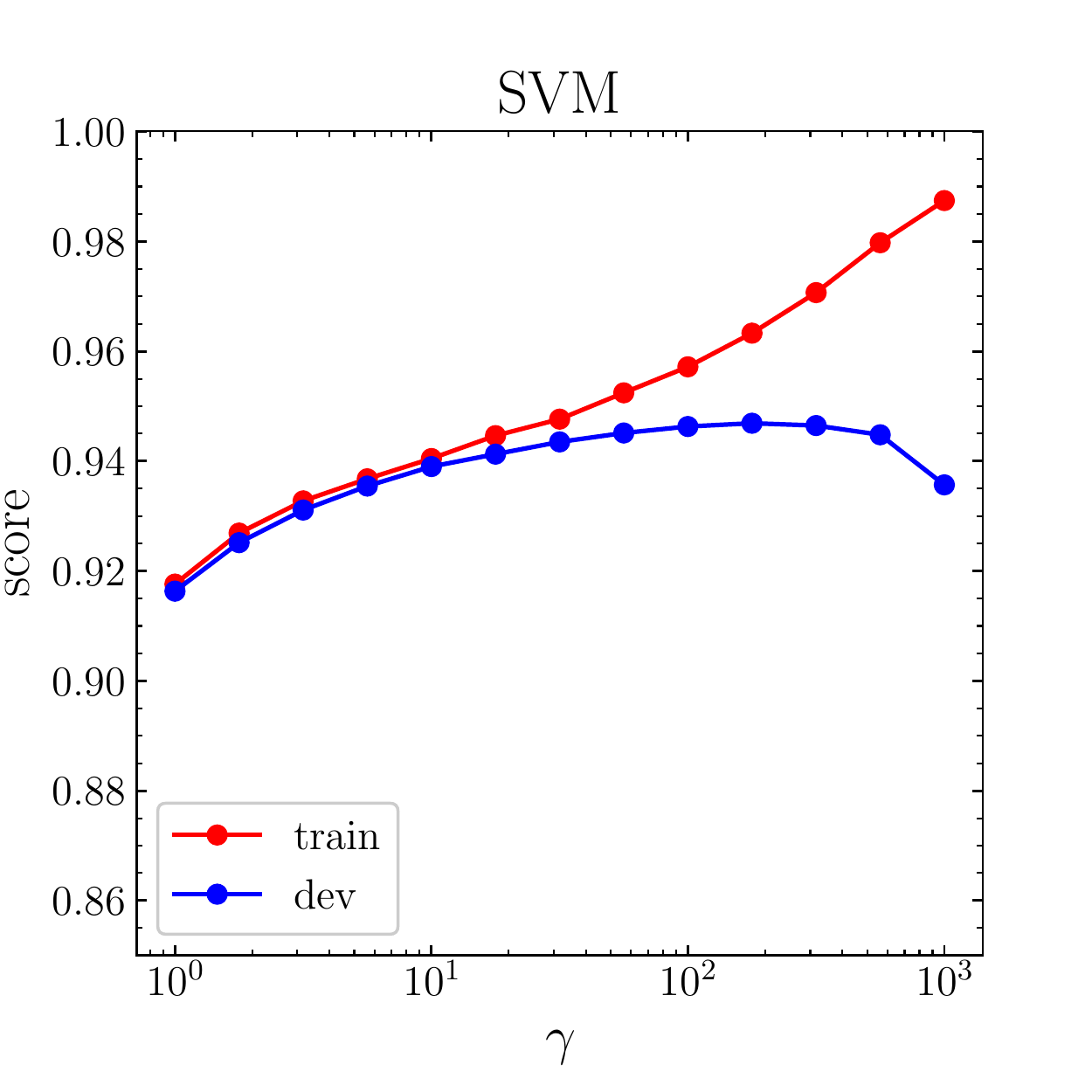}
\end{center}
\caption{
The accuracy of the SVM algorithm on the training and development sets, as a function of the $\gamma$ parameter
defined in Eq.~(\ref{gamma}). 
The best performance of the SVM algorithm on the development set
  is reached at $\gamma \simeq 100$. 
 }
\label{fig:vm}
\end{figure}

 The next algorithm that we consider is the k-nearest neighbours (\textbf{KNN}) algorithm, in which
 the classification is made on the basis of the observation in the k-nearest neighbours. We here chose $k=3$, though we have
 confirmed that the performance does not change once other reasonable values for $k$ are considered.
 We also
 base our decision on the distance from the point. 
 KNN is particularly appropriate for  problems where the data points
 can suitably cover the whole parameter space and the  
observations closest to a given  point should provide us with the most probable observation at that point. 
Though KNN is perhaps the simplest and most straightforward ML algorithm in many different respects,
its implementation in detecting $\nu$ELN crossings in CCSN and NSM simulations may not be that efficient.
This is because at any classification, one needs to load and analyze the whole training dataset, which could be computationally
expensive.

 Apart from LR and KNN, we also tried the performance of support vector machines (\textbf{SVM}s) 
 with the radial basis function (RBF) kernel and decision tree (\textbf{DT}) algorithm. SVM tries to find the best
 hyperplane in the feature space, which separates the different classes in the dataset such that
  the margins are maximised. This means that the distance between the data
 points in different classes and the classifier hyperplane is maximum. The RBF kernel is defined as,
 \begin{equation}\label{gamma}
\mathcal{K}(x,x') = \exp(-\gamma ||x-x'||^2),
\end{equation}
 where $\gamma$ is a hyper-parameter which needs to be determined by the evaluation of the SVM performance in 
 the development set. 
The RBF kernel is based on finding similarities between the test data point and the ones in the training set, where  $\gamma$ 
 determines the  distance up to which the similarities can efficiently make an impact.
 
 As one can see in Fig.~\ref{fig:vm}, the best performance of the SVM algorithm on the development set
  is obtained at $\gamma \simeq 100$. 
  One should also note that the training set  can reach an accuracy of $\sim 100\%$. This should come as no surprise
  since for very large $\gamma$'s, the decision for each point in the training set  is mostly based on the observation at itself.

  DT is another powerful method in classification and prediction, which has a flowchart-like tree structure.
  It include internal nodes and branches, where 
   each internal node denotes a test on a feature and each branch represents the outcome of the test.
   There are also leaf nodes which can be considered as  terminal nodes holding the class labels.

  \begin{table}[tbh!]
\centering
\caption{A summary of the metric scores of our ML algorithms. Note that almost all the tried ML algorithms
can easily reach scores more than $90\%$. In front of each algorithm, one can find its accuracy, i.e., LR, KNN, SVM, and DT have
accuracies of 93\%, 95\%, 95\%, and 94\%, respectively. Though the LR algorithm scores can be less than the others by a  few percents,
its easy implementation and probabilistic interpretation makes it the most promising  candidate ML algorithm to be used in detecting
$\nu$ELN crossings.}
\begin{tabular}[t]{|lcc|c|}
\hline
& \textcolor{black}{ \textbf{{Logistic Regression}} (93\%)}    \\
\hline
& precision & recall & $F_1$-score \\
\hline
no  crossing & 83\% & 93\% & 88\% \\
 crossing&97\%&93\% & 95\% \\
\hline
&\textcolor{black}{  \textbf{{KNN (n=3)}} (95\%)  } \\
\hline
&precision&recall & $F_1$-score \\
\hline
no  crossing&90\%&90\%&90\%\\
 crossing&96\%&96\%&96\%\\
\hline
&\textcolor{black}{   \textbf{{SVM}} (95\%)}\\
\hline
&precision&recall & $F_1$-score\\
\hline
no  crossing&92\%&90\%&91\%\\
 crossing&96\%&97\%&97\%\\
\hline
&\textcolor{black}{   \textbf{{Decision tree}} (94\%) }\\
\hline
&precision&recall & $F_1$-score \\
\hline
no  crossing&89\%&88\%&89\%\\
 crossing&96\%&96\%&96\%\\
\hline
\end{tabular}
\label{tab:tab}
\end{table}%

  The general performances of our ML algorithms are  presented in Table~\ref{tab:tab}.
  As can be  seen, all the ML algorithms  can easily reach accuracies $\gtrsim 90\%$.
  One interesting point  in  Table~\ref{tab:tab} is that
  the precision/recall scores of the no-crossing class is systematically lower 
  for all the ML algorithms. This can be explained given the fact that for our dataset,
  the regions of no-crossing are too close to the boundary region between crossings/no-crossing regions,
  which makes the classification more prone to error.
  This is well illustrated in the middle panel of Fig.~\ref{fig:cross}, where 
the region of the no-crossing class is narrow and therefore many of the data points are too close to the boundary 
of the crossing/no-crossing regions, unlike what one has for the crossing class.
  However, one should bear in mind that this might be also just an artefact of the unrealistic dataset we utilise here and hence
  should be explored in more details once more realistic data are available.
    One should also note that the maximum accuracy is almost  similar   in all different algorithms, namely $\sim 95\%$.
  This is indeed the Bayes error rate of this problem,
  which is the maximum   accuracy  accessible in ideal situations. This  implies that the ML algorithms do
  a fantastic  job in our classification problem.  As a matter of fact, 
   the main source of error in this problem
  is  the noisiness of the target labels. In other words, there are regions in the  $F_{\nu_e} - F_{\bar\nu_e} $ parameter space
  for which  the different parametric angular distributions lead to different results/labels regarding the existence of $\nu$ELN crossings.
  In this problem, 
  one can find the maximum Bayes error rate  by checking the asymptotic accuracy of the knn
  method once the number of data points goes to infinity. We have confirmed that the value of the Bayes error rate
  obtained this way is consistent with the values observed in Table~\ref{tab:tab}, i.e., $\sim 95\%$.

  \begin{figure*}
    \centering
    \subfigure{\includegraphics[width=0.32\textwidth]{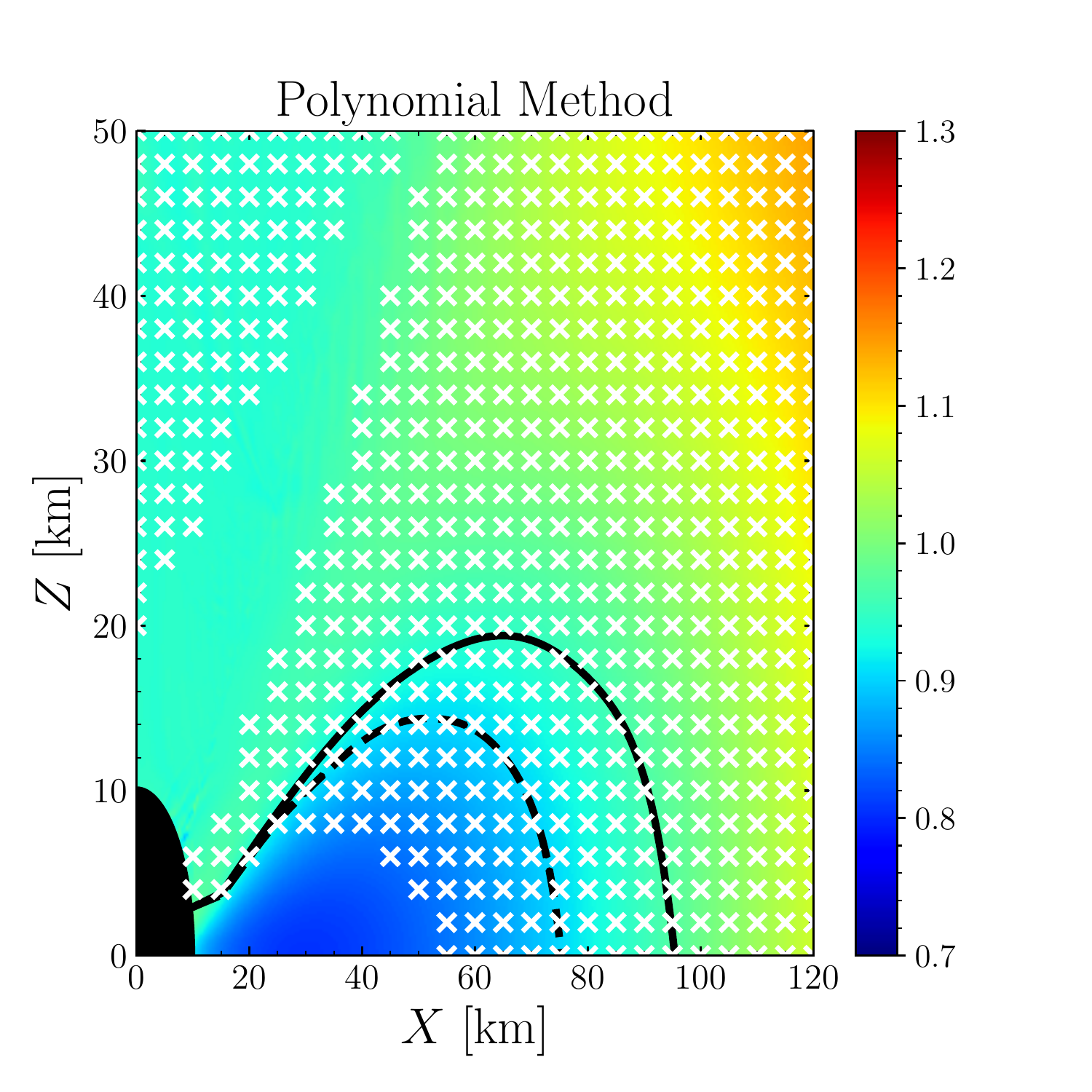}} 
    \subfigure{\includegraphics[width=0.32\textwidth]{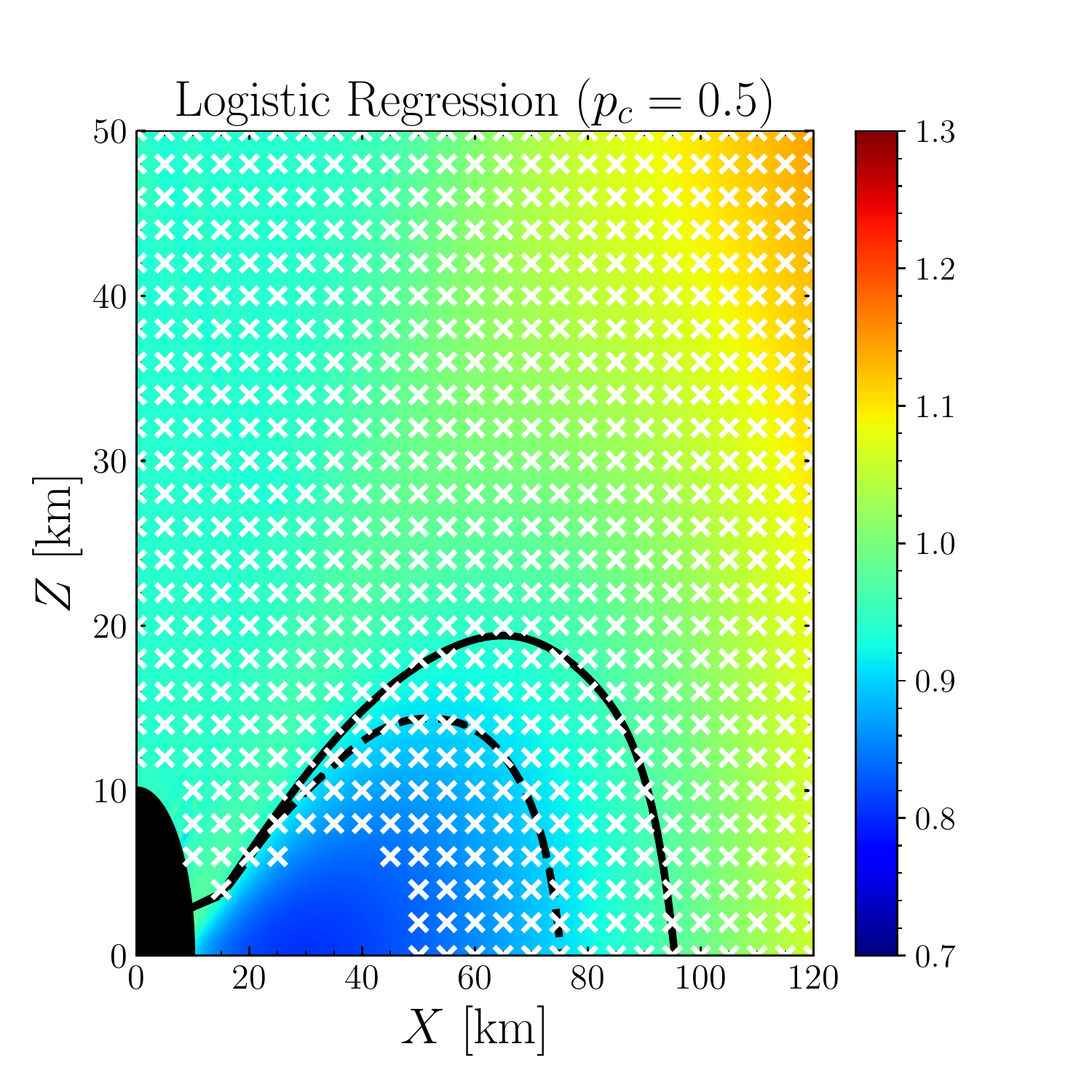}} 
    \subfigure{\includegraphics[width=0.32\textwidth]{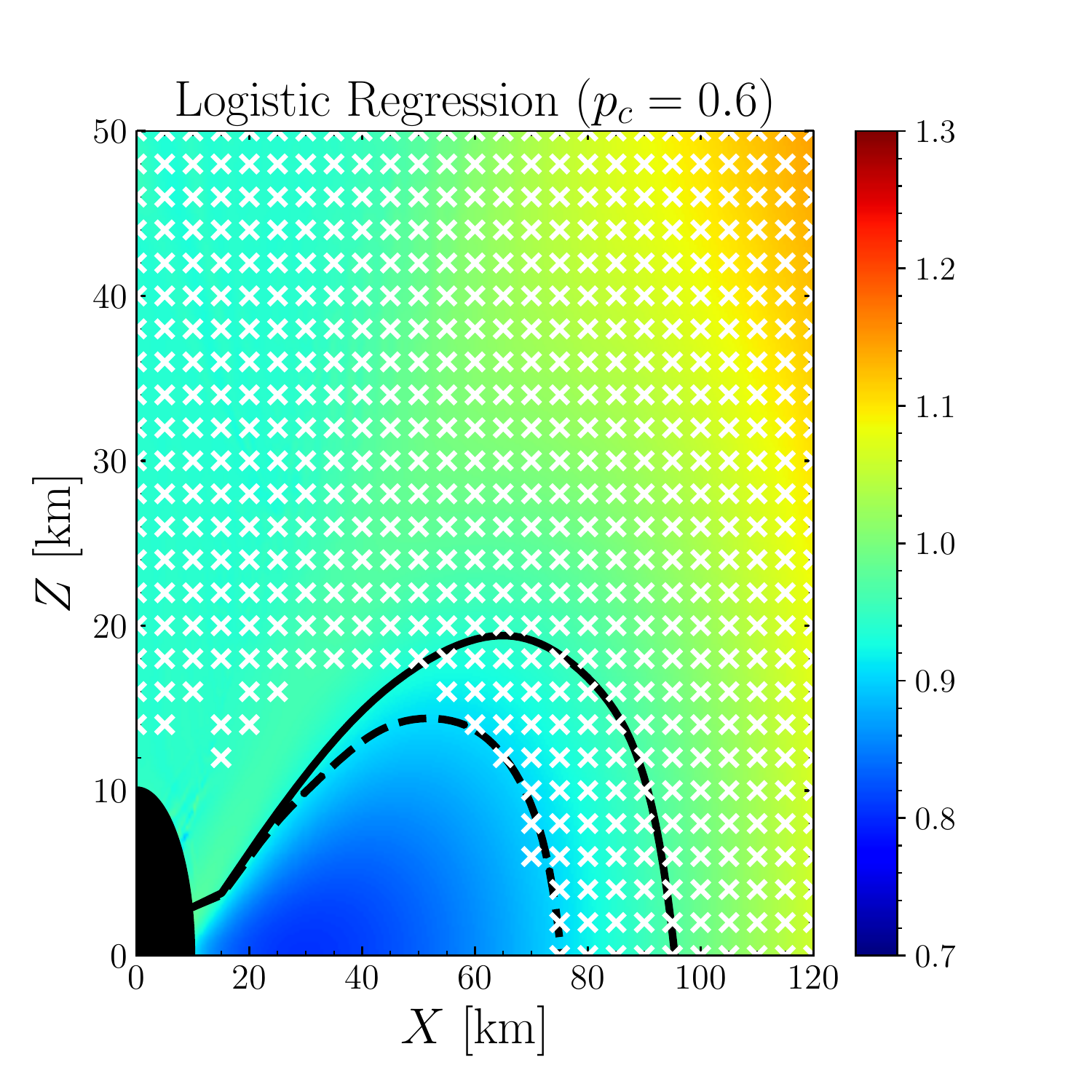}}
    \subfigure{\includegraphics[width=0.32\textwidth]{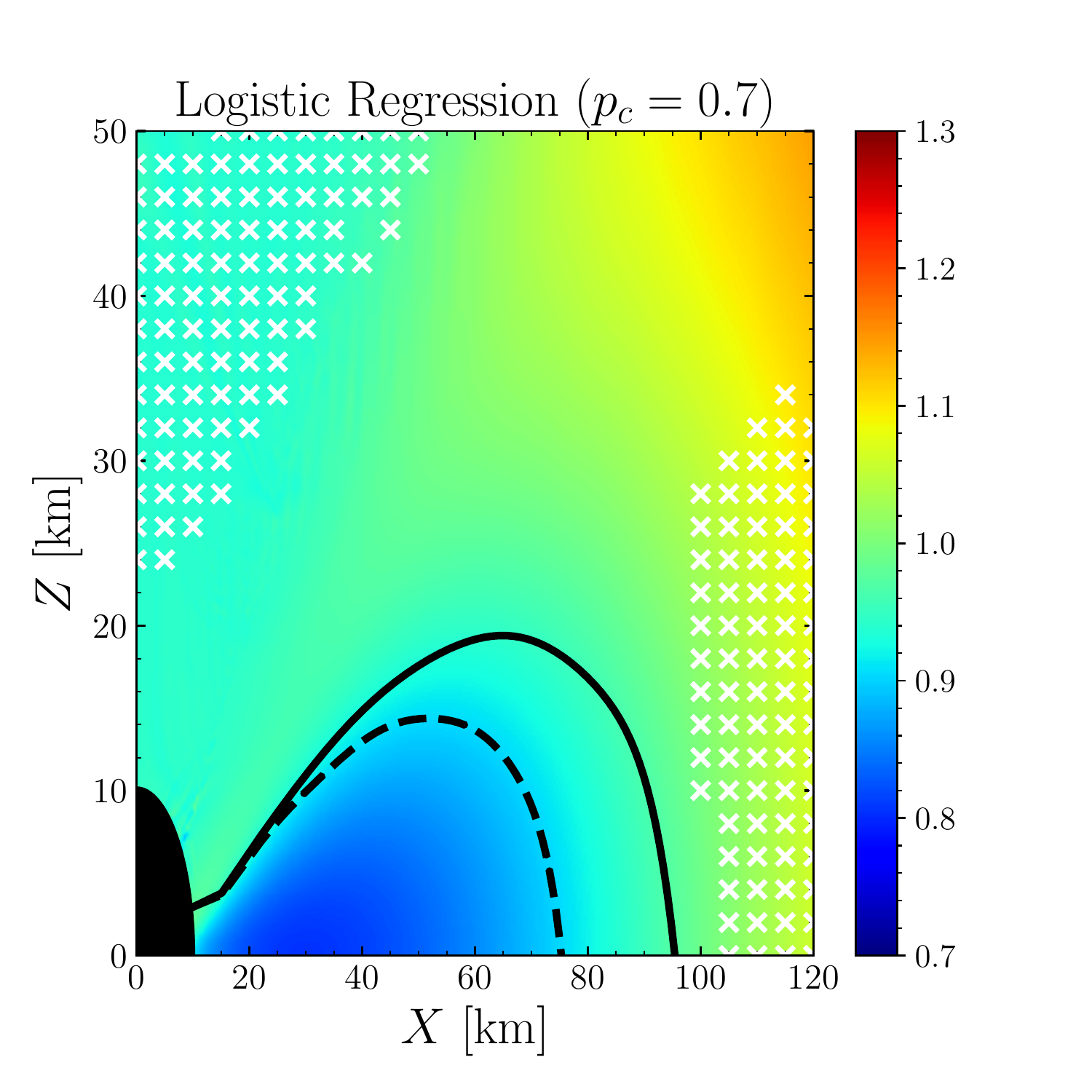}}
     \subfigure{\includegraphics[width=0.32\textwidth]{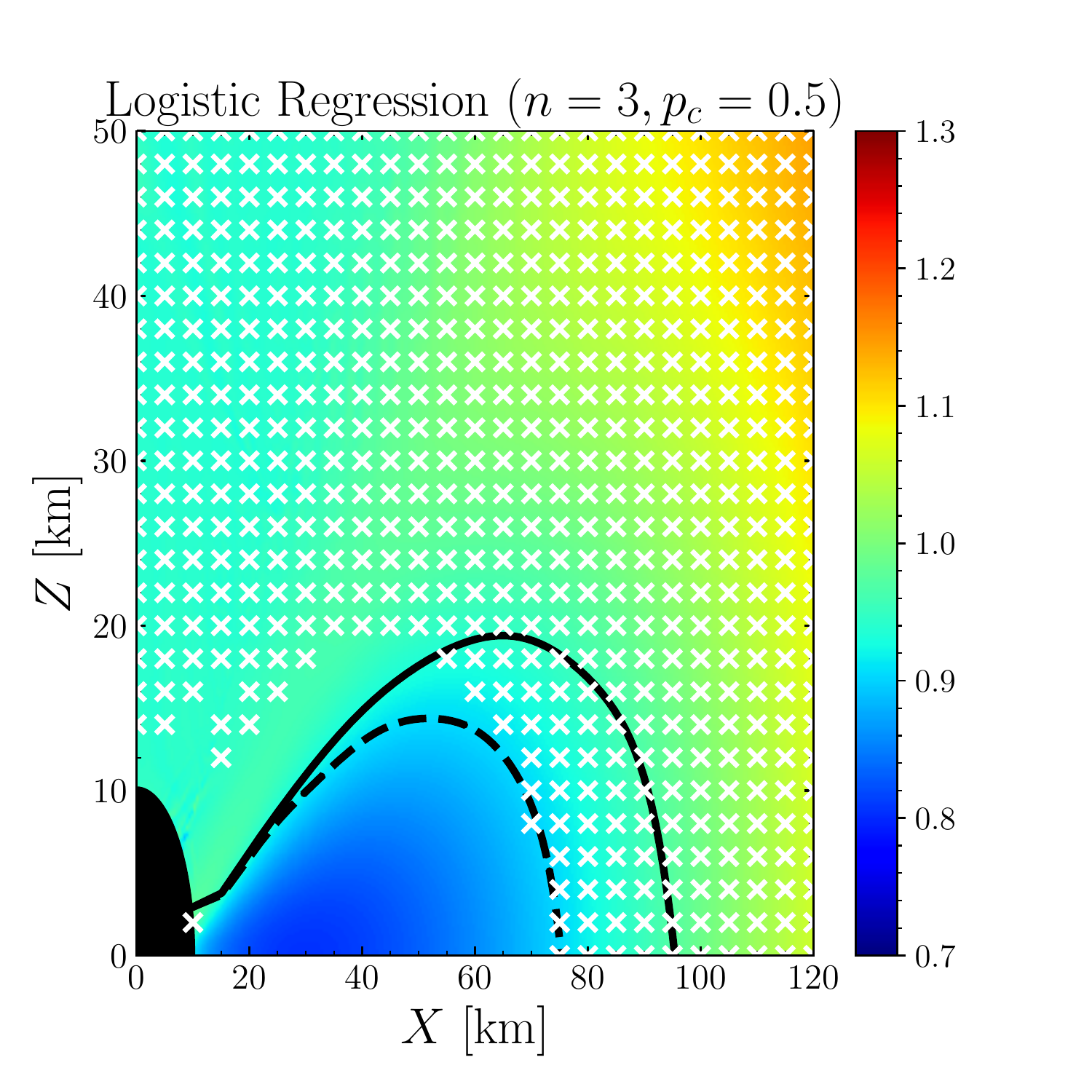}}
     \subfigure{\includegraphics[width=0.32\textwidth]{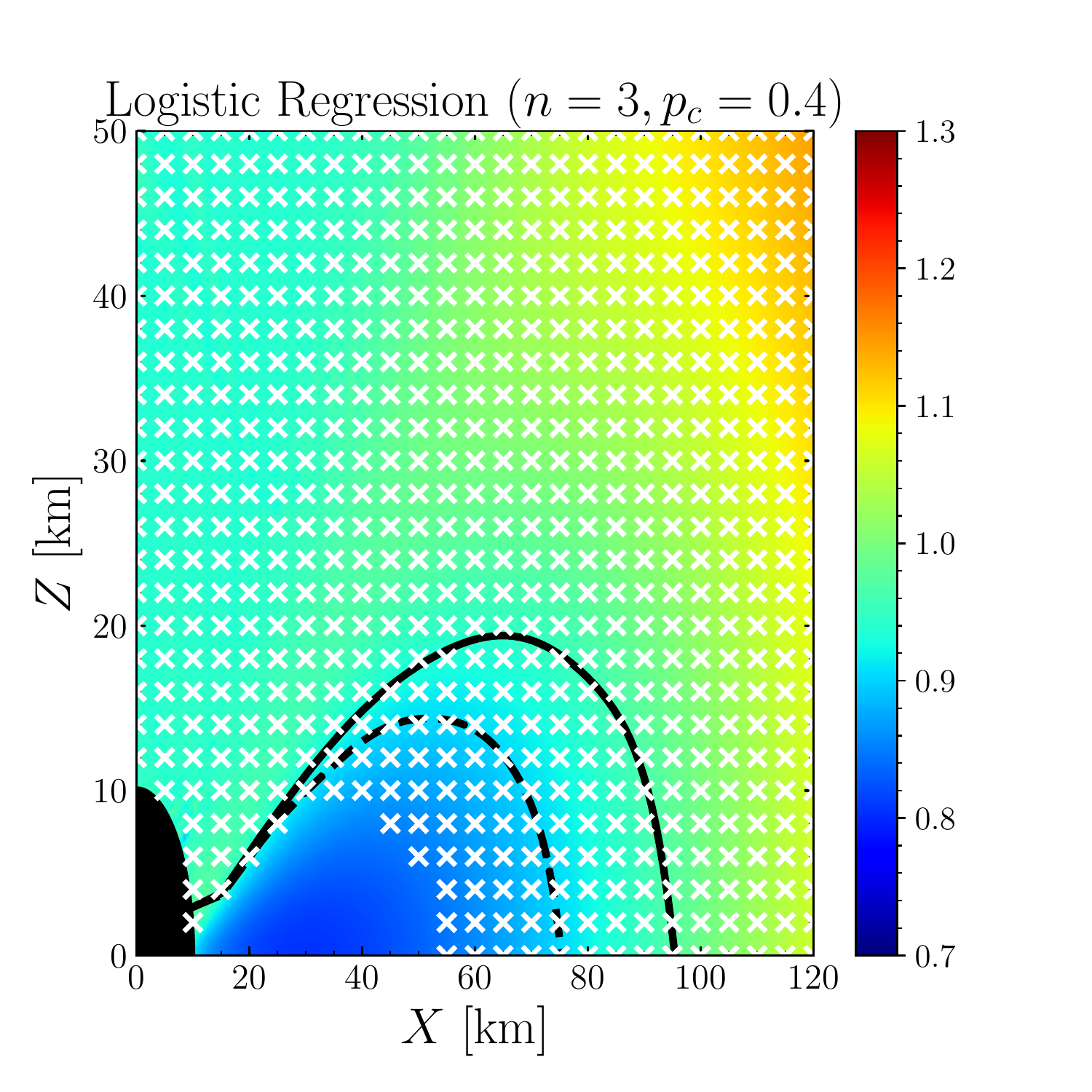}}
    \caption{ $\nu$ELN crossings (white crosses) in the polar plane of the axisymmetric simulation (model M3A8m3a5 from Ref.~\cite{Just:2014fka}) at 
$t=50$ ms. The z axis is the symmetry axis and x  is the cylindrical radius, the color map shows
the $\nu_e - \bar\nu_e$ asymmetry parameter, $\alpha = n_{\bar\nu_e}/n_{\nu_e}$, and
the solid (dashed) lines indicate the locations of the $\nu_e$ ($\bar\nu_e$) neutrinospheres.
   While the upper left panel shows the crossings captured by the polynomial method,
   the rest of the panels indicate the  $\nu$ELN crossings captured by the LR algorithm
    with different threshold densities, $p_c$, and  polynomial degrees ($n=9$ for the upper right and lower left panels and $n=3$
    for the lower middle and right panels).  }
    \label{fig:NSM}
\end{figure*}

  \subsection{Applying ML to a NSM remnant simulation}\label{sec:real}
  Having trained and tested our ML algorithms by the parametric neutrino angular
  distributions, we now test them also on some realistic data obtained from simulations
  of a NSM remnant (model M3A8m3a5 from Ref.~\cite{Just:2014fka}). This way, one can get a flavor of how our ML algorithms perform on datasets which 
  are remarkably different from what they have been trained on.
   This data has been explored in Ref.~\cite{Just2022c}, where
  the polynomial method developed in Ref.~\cite{Abbar:2020fcl} was used to detect  
   $\nu$ELN crossings. As discussed before, the advantage of the polynomial method is in that 
  it is very accurate in what it can detect, i.e., it has a precision of $100\%$,
  though its recall is not guaranteed to be high. 
  
  In Fig.~\ref{fig:NSM}, we present our results on using different methods 
  to detect    $\nu$ELN crossings in
   the simulation of the NSM remnant model studied in Ref.~\cite{Just2022c}, in the time snapshot $t=50$~ms.
    The upper left panel shows the crossings captured by the polynomial method, which is similar to Fig.~1
    of the Ref.~\cite{Just2022c}. The rest of the panels indicate the  $\nu$ELN crossings captured by the LR algorithm
    with different threshold densities, $p_c$, and  polynomial degrees.  
   As one can clearly see,  all the $\nu$ELN crossings captured by the
  polynomial method can be also detected  by the LR
  algorithm with $p_c = 0.5$ (upper middle panel). Besides, LR can detect a number of other  $\nu$ELN crossings, particularly
  in the region where $\alpha\simeq1$. This is, in fact, not surprising. As also discussed in Ref.~\cite{Just2022c},
  the   $\nu$ELN crossings are very likely to occur in this regions. However, the polynomial 
  method can miss these crossings due to its important limitations regarding its dependence on the amount of available neutrino angular
  information, i.e., the number of available moments. But the ML method does not have this limitation and it is not surprising that it captures a number of crossings 
  in this region. 
  There is indeed an important observation here. 
  Once machine learning is considered, the information of the angular distribution shape leaks 
from the data  to the decision maker. So although the higher moments are not available, one has still some information about them coming from the 
patterns in the  data. In contrast, in the polynomial method the information is solely  provided through
the angular moments. 
   In addition, the number of detected crossings
  decreases with increasing $p_c$, as expected (upper right and lower left panels). In all these three panels, the polynomial
  degree is taken to be $n=9$.  

Though the best accuracy of the LR algorithm is reached for $n=9$, the accuracy is more or less acceptable for $n\geq3$.
Specifically, $n=3$ is an interesting case given the fact that it requires much less amount of computations. In the
lower middle and right  panels of  Fig.~\ref{fig:NSM}, the LR classification results with $n=3$ are presented for $p_c=0.5$
and $0.4$, respectively. Though the accuracy of the LR algorithm with $n=3$ is lower than that of $n=9$, it can capture 
almost all the $\nu$ELN crossings detected by the LR algorithm for $n=9$ and $p_c=0.5$,
 once lower $p_c$ is tried (lower right panel). 

We noticed that the accuracy of the  ML algorithms could be remarkably reduced if the $\alpha$ range of the test set is significantly
outside the one of the training set. This implies that to have an accurate ML algorithm, one should have access 
to a very generic training set  including all the relevant values of $\alpha$. This can be indeed difficult given
the scarceness of realistic neutrino angular distributions. This again justifies the application of the parametric angular distributions
to training the ML algorithms.

We have here only discussed the results of the LR algorithm due to its importance in detecting $\nu$ELN crossings. But the other 
algorithms also perform similarly well on this model.
Though  our ML algorithms were trained on some parametric angular distributions, they performed 
remarkably well on some  realistic data they had never seen. We consider this as a very promising sign  of the applicability of
ML in detecting  $\nu$ELN crossings in the most extreme astrophysical settings (models).

\section{DISCUSSION AND OUTLOOK}
We have studied the application of ML to detecting $\nu$ELN crossings in CCSN and NSM models.
Unlike the other existing methods which are either inefficient or very slow, once trained,  ML  
can be very fast and efficient in detecting the $\nu$ELN crossings. Hence, it provides one with a strong 
tool to explore  FFCs in CCSN and NSM simulations on the fly. 
We show that ML algorithms can reach very high accuracies, $\sim 95\%$, which is almost the Bayes error rate of this problem,
though the exact accuracy should depend
on the environment  where the ML algorithms are employed.

For training our ML algorithms, we used two parametric neutrino angular distributions already considered in the literature, namely
the maximum entropy and the Gaussian distributions. Choosing such parametric  distributions 
helps overcome the overfitting, which can be particularly  serious in this problem. To be more specific,
ML classifications can be very sensitive 
 to the accessible values of $\alpha$ in the  training set.

Of particular interest is the LR algorithm which can be easily implemented in CCSN and NSM simulations.
At the same time, the probabilistic   
interpretation of LR allows one to have a strong tool to control/evaluate the
the extent of the impact of FFCs on the   physics of CCSNe and NSMs.

In order to check the efficiency and reliability of our trained ML algorithms, we tested them on some data obtained from a realistic NSM remnant 
simulation. Though the  ML algorithms had never seen such data, they perform remarkably well on them. 
We consider this as a very promising achievement supporting the application of ML to detecting $\nu$ELN crossings in NSM and 
CCSN simulations. Despite this, we would like to emphasise that our study is just meant to introduce this novel idea and more research 
is still required to train more accurate ML algorithms to detecting  $\nu$ELN crossings in realistic NSM and 
CCSN models.

One of the limitations of our study is that our ML algorithms are only sensitive to the crossings in the 
 $\nu$ELN angular distribution once it is integrated over $\phi$. In realistic CCSN and NSM simulations, 
 there could exist $\nu$ELN  crossings in the $\phi$ direction, if the the angular distribution is asymmetric enough in $\phi$.
 For such cases, one should use directly the realistic data since reasonable parametric angular distributions
 do not currently exist. If such data are available, one should also consider a larger number of features in the ML method
 since the $\phi$-related moments can also play a role.
 Another limitation of our research comes from the assumption that 
 the angular distributions of  $\nu_x$ and  $\bar\nu_x$ are the same. In realistic situations, one can
 have $f_{\nu_x} \neq f_{\bar\nu_x}$, particularly due to the creation of muons in the densest regions
 of CCSN and NSM environments~\cite{Bollig:2017lki, Capozzi:2020syn}. 
 This makes it necessary to explore  ML algorithms 
 which are capable of detecting $\nu$ELN crossings once $\nu_x$ and  $\bar\nu_x$ are taken into account separately.
 
 In brief, though our study provides the first step in applying ML techniques to exploring FFCs in CCSN and NSM
simulations and proves their applicability in this business, further research is still required to investigate more realistic and complicated situations. 


\section*{Acknowledgments}
I would like to thank O. Just for providing the data of model
M3A8m3a5, and
 Georg Raffelt, Jakob Ehring, Meng-Ru Wu, and Zewei Xiong 
for useful discussions. 
This work was supported by the German Research Foundation (DFG) through
the Collaborative Research Centre  ``Neutrinos and Dark Matter in Astro-
and Particle Physics (NDM),'' Grant SFB-1258, and under Germany’s
Excellence Strategy through the Cluster of Excellence ORIGINS
EXC-2094-390783311.
I would also like to acknowledge the use of the following softwares: \textsc{Scikit-learn}~\cite{pedregosa2011scikit}, \textsc{Matplotlib}~\cite{Matplotlib}, \textsc{Numpy}~\cite{Numpy}, \textsc{SciPy}~\cite{SciPy}, and \textsc{IPython}~\cite{IPython}.

\bibliographystyle{elsarticle-num}
\bibliography{Biblio}

\clearpage

\end{document}